\documentclass[aps,pre,graphicx,groupedaddress,reprint,showpacs]{revtex4-1}
\usepackage{amsmath}
\usepackage{mathtools}
\usepackage{amssymb}
\usepackage{graphicx}
\usepackage{hyperref}
\usepackage{tensor}
\usepackage{txfontsb} 
\usepackage{multirow}
\DeclareMathAlphabet{\mathcal}{OMS}{cmsy}{m}{n} 
\hypersetup{
    colorlinks=true,
    linkcolor=blue,
    citecolor=blue,
    filecolor=blue,
    urlcolor=blue,
}
\usepackage{listings}

\DeclareSymbolFont{largesymbolsCM}{OMX}{cmex}{m}{n}

\let\sum\relax
\DeclareMathSymbol{\sum}{\mathop}{largesymbolsCM}{"50}
\DeclareSymbolFont{largesymbolsCM}{OMX}{cmex}{m}{n}

\let\prod\relax
\DeclareMathSymbol{\prod}{\mathop}{largesymbolsCM}{"51}
\DeclareSymbolFont{largesymbolsCM}{OMX}{cmex}{m}{n}

\let\intop\relax
\DeclareMathSymbol{\intop}{\mathop}{largesymbolsCM}{"52}

\let\int\relax
\def\int{\intop\nolimits}

\makeatletter
\renewcommand*\env@matrix[1][\arraystretch]{%
  \edef\arraystretch{#1}%
  \hskip -\arraycolsep
  \let\@ifnextchar\new@ifnextchar
  \array{*\c@MaxMatrixCols c}}
\makeatother

\begin{document}

\title{ \ \\ Linear embedding of free energy minimization}

\author{Jonathan E. Moussa}
\email{godotalgorithm@gmail.com}
\affiliation{Center for Computing Research, Sandia National Laboratories, Albuquerque, New Mexico 87185, USA}

\begin{abstract}  \begin{minipage}{0.79\textwidth}
\ \\

\ \ \ Exact free energy minimization is a convex optimization problem that is usually approximated with stochastic sampling methods.
Deterministic approximations have been less successful because many desirable properties have been difficult to attain.
Such properties include the preservation of convexity, lower bounds on free energy, and applicability to systems without subsystem structure.
We satisfy all of these properties by embedding free energy minimization into a linear program over energy-resolved expectation values.
Numerical results on small systems are encouraging, but a lack of size consistency necessitates further development for large systems.


\end{minipage}\end{abstract}

\maketitle

Deterministic approximations of free energy minimization trace back to the early work of Bethe \cite{Bethe}.
His approach was later generalized into the cluster variation method (CVM) \cite{Kikuchi}
 and has served as the foundation for a broad class of modern statistical inference methods \cite{graphical_models}.
Such deterministic methods have not been as successful as stochastic methods such as the Metropolis algorithm \cite{Metropolis}
 because of a few persistent technical problems.
Relaxation of the convex free energy minimization problem should produce a lower bound on free energy that is the unique global minimum of a simpler convex optimization problem.
Only recently have methods begun to achieve this goal by restricting statistical correlations to tree \cite{tree_entropy} or causal \cite{causal_entropy} structures.
A more persistent problem is that systems must be decomposable into subsystems with a moderate number of microstates.
This is a problem for rotors in classical systems or bosons in quantum systems where all subsystems have an infinite number of microstates without artificial truncations.

At zero temperature, the minimization of free energy over microstate probability distributions reduces to a search for a microstate with the lowest energy.
If there is a large number of microstates, then this is still a hard optimization problem.
However, relaxation methods in discrete optimization such as the cutting-plane method \cite{cutting_plane}
 can reduce hard optimizations to computationally tractable convex optimizations.
The relaxed problems optimize over supersets of the original domain and produce lower bounds on minima.
While there are no known universal methods for relaxing all hard problems,
 a variety of problem-specific relaxation methods exist \cite{relaxation_survey}.

In this paper, we study a convex relaxation of free energy minimization to linear programs.
The primary difficulty with applying existing relaxation techniques to this problem is the nonlinearity of entropy on microstate probabilities.
All known relaxations depend on linear information inequalities \cite{info_inequal} and a decomposition of systems into small subsystems.
We derive an alternate approach that embeds each expectation value in a distribution of contributions over microstate energy.
This is a generalization of the energy density of states sampled by the Wang-Landau algorithm \cite{Wang_Landau}.
The embedding linearizes the free energy, which enables a cutting-plane relaxation.

Our goal is to compute the minimum free energy $F(T)$
 at a temperature $T$ of a system with $N$ microstates,
\begin{align} \label{free_energy}
 & F(T) = \min_{p_i} \sum_{i=1}^N \left( E_i p_i + T p_i \ln p_i \right) \notag \\
 & \mathrm{subject \ to} \ \sum_{i=1}^N p_i = 1 \ \mathrm{and} \ p_i \ge 0 .
\end{align}
Microstate $i$ has energy $E_i$ and probability $p_i$.
$T$ is in units of energy.
The minimizer $p_i(T)$ of Eq. (\ref{free_energy}) is the
 Boltzmann distribution with partition function $Z(T)$,
\begin{equation} \label{minimizer}
 p_i(T) = \frac{\exp(- E_i / T)}{Z(T)} , \ Z(T) = \sum_{i=1}^N \exp(- E_i / T ) .
\end{equation}
$F(T)$ is not efficient to compute directly when $N$ is large.

We assume a generic form of structure in Eq. (\ref{free_energy}) that does not make reference to subsystems.
The assumption is that a system can be described
 by the expectation values of a small number of observables, $n \ll N$.
We can then reduce $p_i$ to $n$ expectation values $x_i$ and $E_i$ to $n$ coefficients $h_i$ as
\begin{equation} \label{expectation}
 x_i = \sum_{j=1}^N B_{i,j} p_j \ \mathrm{and} \ \sum_{i=1}^n h_i B_{i,j} = E_j
\end{equation}
 where observables are defined by a matrix $B_{i,j}$ with linearly independent columns and $B_{1,i} = 1$.
A standard example is a description of spin systems by few-spin expectation values.

To simplify Eq. (\ref{free_energy}), we must enforce $p_i \ge 0$ with only $x_i$.
The feasible set of $x_i$ values is hard to compute.
Instead, we relax the $N$ constraints on $p_i$ to $m \ll N$ constraints on $x_i$ as
\begin{equation} \label{relax}
 p_i \ge 0 \ \mathrm{and} \ \sum_{j=1}^n  A_{i,j} B_{j,k} \ge 0 \ \Rightarrow \ \sum_{j=1}^n A_{i,j} x_j \ge 0 ,
\end{equation}
 for an $m$-by-$n$ matrix $A_{i,j}$ that identifies linear combinations of observables with guaranteed nonnegative expectation values.
Eq. (\ref{relax}) is the central approximation of our $F(T)$ lower bound construction.
A standard example is nonnegativity of marginal probabilities for few-spin subsystems of a spin system.

In the $T = 0$ limit, $F(T)$ reduces to finding the microstate $i$ that has the lowest energy $E_i$.
This is a linear program in $N$ dimensions with variables $p_i$.
We relax Eq. (\ref{free_energy}) with Eq. (\ref{relax}) to a linear program in $n$ dimensions with variables $x_i$,
\begin{align} \label{free_relax}
 & F(0) \ge \min_{x_i} \sum_{i=1}^n h_i x_i \notag \\
 & \mathrm{subject \ to} \ x_1 = 1 \ \mathrm{and} \ \sum_{j=1}^n A_{i,j} x_j \ge 0 .
\end{align}
In this context, Eq. (\ref{relax}) is a form of the cutting-plane
 method \cite{cutting_plane} in discrete optimization that simplifies the feasible set.

For $T > 0$, the convex relaxation of $F(T)$ in Eq. (\ref{free_energy}) is more
 difficult because of the nonlinearity of free energy in $p_i$, which is hard to approximate with $x_i$.
Inspired by the Wang-Landau algorithm \cite{Wang_Landau}, 
 we decompose $p_i$ into an integral over energy density of states.
We write this using normalized distributions $P_i(s)$
 over the surprisal, $s = \ln Z(T) + E/T$, such that
\begin{equation} \label{spectral_rep}
  \int_{0}^{\infty} P_i(s) \, ds = 1 \ \mathrm{and} \ \int_{0}^{\infty} P_i(s) \, e^{-s} ds = p_i .
\end{equation}
The $N$ distributions $P_i(s)$ solve a continuous linear program
\begin{align} \label{free_relax2}
 & F(T) = \min_{P_i(s)} \sum_{i=1}^N \int_{0}^{\infty} (E_i - T s) P_i(s) \, e^{-s} ds \notag \\
 & \mathrm{subject \ to} \ \sum_{i=1}^N \int_{0}^{\infty} P_i(s) \, e^{-s} ds = 1 ,  \ P_i(s) \ge 0, \notag \\
 & \ \ \ \ \ \ \ \ \ \ \ \ \ \ \ \ \mathrm{and} \ \int_{0}^{\infty} P_i(s) \, ds = 1.
\end{align}
This is a straightforward linear embedding of Eq. (\ref{free_energy}).
Every normalized nonnegative distribution collapses and results in a
 Dirac delta function, $P_i(s) = \delta(s - \ln Z(T) - E_i/T)$.

The $T=0$ relaxation method can now be applied at $T>0$.
First, we relax the distributional nonnegativity of $P_i(s)$,
\begin{equation} \label{relax2}
 P_i(s) \ge 0 \ \Rightarrow \ \sum_{j=1}^n A_{i,j} X_j(s) \ge 0 .
\end{equation}
Distributions $X_i(s)$ subsume $x_i$ and Eq. (\ref{relax2}) subsumes Eq. (\ref{relax}).
Next, we relax Eq. (\ref{free_relax2}) to the continuous linear program
\begin{align} \label{free_relax3}
 & F(T) \ge \min_{X_i(s)} \sum_{i=1}^n \int_{0}^{\infty} (h_i - \delta_{1,i} T s) X_i(s) \, e^{-s} ds \notag \\
 & \mathrm{subject \ to} \ \int_{0}^{\infty} X_1(s) \, e^{-s} ds = 1 ,  \ \sum_{j=1}^n A_{i,j} X_j(s) \ge 0, \notag \\
 & \ \ \ \ \ \ \ \ \ \ \ \ \ \ \ \ \mathrm{and} \ \int_{0}^{\infty} X_i(s) \, ds = g_i := \sum_{j=1}^N B_{i,j} .
\end{align}
The number of optimization variables has been reduced from $N$ to $n$ distributions.
This bound is equivalent to Eq. (\ref{free_relax}) in the $T = 0$ limit.
Further relaxation requires discretization.

We finally relax Eq. (\ref{free_relax2}) to a finite linear program by testing
 nonnegativity against $r$ nonnegative test functions $t_i(s)$,
\begin{equation} \label{test_functions}
 x_{i,j} = \int_0^\infty X_i(s) t_j(s) \, ds ,
\end{equation}
 that fit all integrands over $s$ with coefficients $u_i$, $v_i$, and $w_i$,
\begin{equation} \label{interpolation}
 \sum_{i=1}^r  \begin{bmatrix} u_i \\ v_i \\ w_i \end{bmatrix} t_i(s) = \begin{bmatrix} 1 \\ e^{-s} \\ s e^{-s} \end{bmatrix} ,
\end{equation}
The relaxed finite linear program with $nr$ variables $x_{i,j}$ is
\begin{align} \label{free_relax4}
 & F(T) \ge \min_{x_{i,j}} \sum_{i=1}^n h_i x_i - T \sum_{i=1}^r w_i x_{1,i} \notag \\
 & \mathrm{subject \ to} \ x_1 = 1 ,  \ \sum_{j=1}^n A_{i,j} x_{j,k} \ge 0, \notag \\
 & \ \ \ \ \ \ \ \ \ \ \ \ \ \ \ \ \mathrm{and} \ \sum_{j=1}^r \begin{bmatrix} u_j \\ v_j \end{bmatrix} x_{i,j} = \begin{bmatrix} g_i \\ x_i \end{bmatrix} .
\end{align}
This naturally reduces to Eq. (\ref{free_relax}) at $T=0$.
The cost of $T>0$ is an increase in the number of variables from $n$ to $nr$.

For the example considered in this paper, we construct test functions as piecewise linear combinations of $1$, $e^{-s}$, and $s e^{-s}$ specifically for odd $r$.
They are defined by a set of $(r+1)/2$ interpolation points $s_i$ and split into two types of functions,
\begin{align}
  t_{2i-1}(p) &= \left\{ \begin{array}{ll} \frac{(1 - s_{i-1} + s) \exp(s_{i-1} - s) - 1}{(1 - s_{i-1} + s_i) \exp(s_{i-1} - s_i) - 1} , & s_{i-1} \le s \le s_i \\
                           \frac{(1 - s_{i+1} + s) \exp(s_{i+1} - s) - 1}{(1 - s_{i+1} + s_i) \exp(s_{i+1} - s_i) - 1} , & s_i \le s \le s_{i+1} \\ 0 , & \mathrm{otherwise}  \end{array} \right. , \notag \\
  t_{2i}(s) &= \left\{ \begin{array}{ll} f_i(s) / f_i(\bar{s}_i) , & s_i \le s \le s_{i+1} \\ 0 , & \mathrm{otherwise}  \end{array} \right. , \notag \\
  f_i(s) &=  \tfrac{s_{i+1}-s}{s_{i+1}-s_i} e^{s_i - s}  + \tfrac{s-s_i}{s_{i+1}-s_i} e^{s_{i+1} - s} - 1, \notag \\
  \bar{s}_i &= 1 + \tfrac{s_i - s_{i+1} \exp(s_i - s_{i+1})}{1 - \exp(s_i - s_{i+1})} ,
\end{align}
 with $s_i < \bar{s}_i < s_{i+1}$.
An arbitrary function $C(s)$ is interpolated by a linear combination of these functions with coefficients
\begin{align}
 c_{2i-1} &= C(s_i) , \notag \\
 c_{2i} &= C(\bar{s}_i) - c_{2i-1} t_{2i-1}(\bar{s}_i) - c_{2i+1} t_{2i+1}(\bar{s}_i) .
\end{align}
This interpolation is exact for $1$, $e^{-s}$, and $s e^{-s}$, and it defines the coefficients in Eq. (\ref{interpolation}).
We use uniform grids of $s_i$ values that cover the range of possible $s$ values in each calculation.

To demonstrate the dimensional reduction capability of our linear embedding of free energy minimization,
 we consider a clock with $N$ microstates labeled by an angle $\theta_i = 2 \pi (i-1) / N$.
Convenient observables of the clock are $n$ moments,
\begin{equation} \label{moments}
 B_{2 i + 1,j} = \cos i \theta_j \ \mathrm{and} \ B_{2 i , j} =  \sin i \theta_j .
\end{equation}
To constraint their expectation values, we must identify linear combinations of moments that are nonnegative for every $\theta_i$.
A clock with moment observables and constraints is an example of non-subsystem structure described by Eqs. (\ref{expectation}) and (\ref{relax}).

We consider a simple clock Hamiltonian with two nonzero terms, $h_2 = -\sin \phi$ and $h_3 = -\cos \phi$.
For ample dimensional reduction opportunities, we consider $N = 2^{20}$.
We solve all of the linear programs with \textsc{ecos} \cite{ECOS}.
Eq. (\ref{free_relax}) is implemented as shown, but Eq. (\ref{free_relax4}) is rescaled to improve numerical stability \cite{supplement}.
We use $v_j x_{i,j}$ as variables instead of $x_{i,j}$, and equations with $g_i = N \delta_{1,i}$ are rescaled by $v_r$.
This avoids large quantities in the program and its solution, but persistent solver stability problems still limit our numerical experiments.

\begin{figure}[b]
\includegraphics{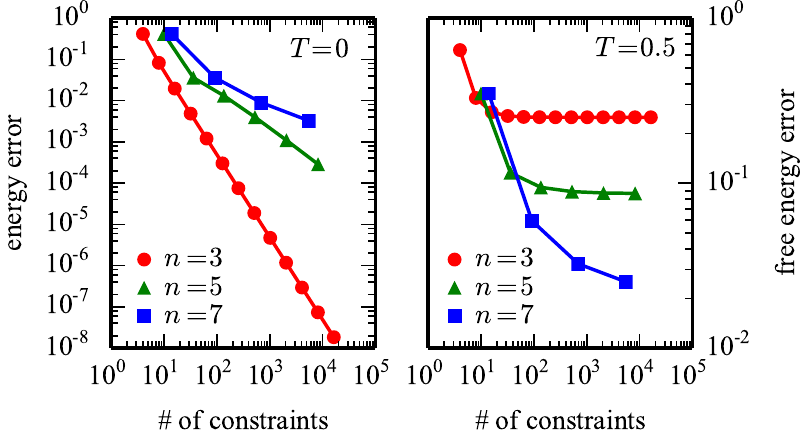}
\caption{\label{fig_clock} (color online) Relaxed energy minimization (left) and free energy minimization (right) for a clock with $N = 2^{20}$ and microstate energies $E(\theta) = -\cos(\theta - \phi)$.
Its exact energies are $F(0) = -1$ and $F(0.5) \approx -7.343$.
At $T = 0$, the error converges with only $n = 3$ observables, and extra observables simply reduce the approximation efficiency.
At $T = 0.5$, the error stagnates with $n = 3$ observables, and extra observables reduce the minimum attainable error.
}
\end{figure}

At $T=0$, it is possible to achieve a tight bound with Eq. (\ref{free_relax}) using only observables that appear in the Hamiltonian with a sufficient set of constraints.
For our example with $x_1$, $x_2$, and $x_3$, these are $N$ constraints defined by
\begin{equation} \label{3obs_constraint}
 A_{i,1} = 1 , \ A_{i,2} = - \sin \theta_i , \ \mathrm{and} \ A_{i,3} = - \cos \theta_i .
\end{equation}
With $N \gg m$, we approximate these constraints by uniformly sampling $m$ points from $\theta_i$.
Error versus $m$ is shown in Fig. \ref{fig_clock}
 with the Hamiltonian tuned to $\phi = \pi/m$ to maximize the error.
There is an $m^{-2}$ scaling of error, which is comparable to the error in estimating
 the minimum of a smooth function from its evaluation at a uniform set of points without interpolation.

AT $T>0$, we set $s_1$ and $s_r$ to be the exact minimum and maximum surprisal values and achieve apparent convergence with $r = 41$.
With $3$ observables, the error saturates in Fig. \ref{fig_clock}.
We extend these calculations to $5$ and $7$ observables with the addition of constraints defined by nonnegative functions of $\theta_i$,
\begin{equation}
 \prod_{j=1}^{q} [1 - \cos(\theta_i - \theta_{\Lambda(j)})] \ge 0,
\end{equation}
 defined by $q$ angles, $\theta_{\Lambda(j)}$, for $q \le (n-1)/2$.
We again limit the number of constraints by uniformly sampling these angles from $\theta_i$.
A systematic reduction of error with increasing $n$ is evident in Fig. \ref{fig_clock} at $T>0$.
Unlike $T=0$, we need to consider observables that do not appear in the Hamiltonian to improve the accuracy at $T>0$.
The cause of these differences between $T=0$ and $T>0$ is unknown and left for future studies.

\newpage

For a many-body method to be accurate for large systems, size-consistency errors must be small \cite{size_consistency}.
These errors occur when separate and combined calculations of two independent subsystems disagree.
We combine Eq. (\ref{free_relax3}) for two subsystems `$L$' and `$R$' by merging the linear programs in a block matrix notation
 to partition observables (`$1$' separated from `$>1$'),
\begin{align} \label{combined_solution}
  \mathbf{A} &=  \begin{bmatrix}[1.1] \mathbf{A}_1^{L} & \mathbf{A}_{>1}^{L} & 0 \\ \mathbf{A}_1^{R} & 0 & \mathbf{A}_{>1}^{R} \end{bmatrix} , \
  \mathbf{B} = \begin{bmatrix}[1.1]  \mathbf{B}_{1 \; \, \, } ^{L} \otimes \mathbf{B}_{1 \; \, \, }^{R} \\ \mathbf{B}_{>1}^{L} \otimes \mathbf{B}_{1 \; \, \, }^{R} \\ \mathbf{B}_{1 \; \, \, }^{L} \otimes \mathbf{B}_{>1}^{R} \end{bmatrix} , \notag \\
  \mathbf{h} &=  \begin{bmatrix}[1.1] h_1^{L} + h_1^{R} \\ \mathbf{h}_{>1}^{L} \\ \mathbf{h}_{>1}^{R} \end{bmatrix} , \
 \mathbf{X}(E) = \begin{bmatrix}[1.1] X_{1 \; \; \, }^{L} \ast X_{1 \; \; \, }^{R} \\ \mathbf{X}_{>1}^{L} \ast X_{1 \; \; \, }^{R} \\ X_{1 \; \; \, }^{L} \ast \mathbf{X}_{>1}^{R} \end{bmatrix} (s) , \notag \\
 (f \ast g) (s) &= \int_{0}^{s} f(s - s') g(s') \, ds' .
\end{align}
By construction, the combined solution is feasible and gives a free energy bound equal to the sum of the subsystem free energy bounds.
However, it is not necessarily a stable solution to Eq. (\ref{free_relax3}) and any further reductions of the free energy bound after reoptimization constitute a size-consistency error.

We study the size-consistency errors of a linear embedding for independent spins in a uniform field.
We model each spin exactly with two observables and two constraints,
\begin{equation}
 2 \mathbf{A} = \mathbf{B} = \begin{bmatrix}[1.1] 1 & 1 \\ 1 & -1 \end{bmatrix} , \
 \mathbf{h} = \begin{bmatrix}[1.1] 0 \\ h_2 \end{bmatrix} .
\end{equation}
This system is invariant to permutations of spins, and we can reduce Eq. (\ref{free_relax3}) for $k$ spins to a single $X_1(s)$ and $X_2(s)$,
\begin{align}
 & F(T) \ge \min_{X_2(s)} \int_{0}^{\infty} \left[ k h_2 X_2(s) - T s | X_2(s)| \right] \, e^{-s} ds \notag \\
 & \mathrm{subject \ to} \ \int_{0}^{\infty} \begin{bmatrix} | X_2(s) | \\ | X_2(s) | e^{-s} \\ X_2(s) \end{bmatrix} \, ds =  \begin{bmatrix} 2^k \\ 1 \\ 0 \end{bmatrix} ,
\end{align}
 with $X_1(s) = |X_2(s)|$ set by saturating the two inequalities.
The optimal $X_1(s)$ reduces to a sum of two Dirac delta functions,
\begin{align}
 X_1(s) &= \sum_{i=0}^1 2^{k-1} \delta(s + F(T)/T + (2 i - 1) k h_2 / T) , \notag \\
 F(T) &= - k T ( \ln 2 + k^{-1} \ln \cosh k h_2 / T ) .
\end{align}
By contrast, the exact $X_1(s)$ has $k+1$ Dirac delta functions and the exact $F(T)$ is a strictly linear function of $k$,
\begin{align}
 X_1(s) &= \sum_{i=0}^k \frac{k!}{i! (k-i)!} \delta(s + F(T)/T + (2 i - k) h_2/T) , \notag \\
 F(T) &= -  k T ( \ln 2 + \ln \cosh h_2 / T ) .
\end{align}
In the $k \rightarrow \infty$ limit, the approximate $F(T) / k$ asymptotes to
\begin{equation}
 F(T)/k \rightarrow - T \ln 2 - h_2 .
\end{equation}
This is accurate in both the $T = 0$ and $T \rightarrow \infty$ limits, but it has a maximum relative error of $\approx 0.5$ at $h_2 / T \approx 0.8$.

In its present form, linear embedding has properties that are complementary to conventional cluster-based methods such as CVM \cite{Kikuchi}.
By construction, cluster expansions of entropy have no size-consistency errors.
However, they are based on exact entropy calculations of small subsystems,
 which assumes the efficient summation over subsystem microstates.
For systems composed of subsystems with a large or even infinite number of microstates,
 brute force cluster expansions are intractable.
Examples include clocks with large $N$ and continuous rotors in the $N \rightarrow \infty$ limit of a clock.
It may be possible to combine linear embedding techniques that reduce the size and cost of
 subsystem calculations with cluster expansions that maintain overall size consistency.
The further development of efficient but non-size-consistent subsystem methods will complement
 orthogonal development of cluster-based methods such as the identification of new information inequalities \cite{info_inequal}.

For linear embedding to develop into a general-purpose free energy minimization method, size-consistency errors must be eliminated.
The simplest possibility would be to identify new constraints that guarantee the stability of combined solutions in Eq. (\ref{combined_solution}).
In addition to the structure postulated in Eqs. (\ref{expectation}) and (\ref{relax}),
 we might attempt to impose predetermined subsystem structure.
When combining calculations of two systems, their subsystem structure must also be merged.
Another plausible size-consistent extension of linear embedding is to extend the $s$ parameter used in Eq. (\ref{spectral_rep})
 into a separate $s_i$ parameter for each microstate that extends the temperature dependence into
 a full characterization of the thermodynamic state space \cite{thermodynamic_length}.
This simplifies the combination of independent subsystems in Eq. (\ref{combined_solution}) from a convolution to a product.
However, it creates a high-dimensional space of variables that must be reduced to
 coincide with the small number of important observables.
A Fisher information matrix can be defined on this space,
 and its spectral nonnegativity might be a useful constraint.
The large number of variables introduced by this extension would need to be compensated for by a large number of constraints.

Linear embedding of free energy minimization preserves its convexity but changes its basic structure.
The exact problem has a linear objective function only at zero temperature,
 while the approximate problem always has a linear objective.
These linear problems are optimized on a boundary of their feasible set.
At finite temperature, the $p_i \ln p_i$ entropy terms in Eq. (\ref{free_energy})
 drive solutions away from the boundary, which are defined by $p_i = 0$ for one or more $i$.
This fundamentally changes how the problem responds to perturbations.
Responses are determined by derivatives of the objective in the nonlinear case
 and by the shape of the boundary in the linear case.
The consequences of this difference are unclear.
Any convenience gained from the use of linear program solvers for free energy minimization is
 possibly offset by a fundamental distortion of the problem.
It is still possible for nonlinear embedding to retain the property
 of having extrema only on the interior of the feasible set.

There are also computational complexity differences in free energy minimization between zero and finite temperature.
At zero temperature, the search for the lowest-energy microstate is NP-hard.
In general, accurate approximation of a solution is intractable on a classical computer in polynomial time \cite{PCP_theorem}.
At finite temperature, free energy approximation is equivalent to approximation of \#P-hard counting problems \cite{entropy_counting}.
While a complete understanding of the relative complexity between zero and finite temperature is not yet known,
 it is possible to approximate \#P problems with access to an NP oracle \cite{sharpP_approx}.

Ultimately, this paper is meant to motivate and inspire new
 ideas in deterministic methods for free energy minimization.
We have developed a linear embedding concept,
 computed a numerical example,
 presented an analytically solvable model,
 and identified the problem of size-consistency errors.
While it is of limited applicability in its present form,
 there are many possible directions of future development.
In particular, there are multiple possible ways to combine linear embedding with
 cluster-based methods to fix size-consistency errors.
Finally, we note that linear embedding has a straightforward quantum generalization
 as do many cluster-based methods \cite{causal_entropy}.
We need methods such as linear embedding to simulate the unbounded Hilbert spaces of bosons without artificial truncations.

\ \\ I thank Norm Tubman for useful discussions.
This work was supported by the Laboratory Directed Research and Development program at Sandia National Laboratories.
Sandia National Laboratories is a multi-program laboratory managed and  
operated by Sandia Corporation, a wholly owned subsidiary of Lockheed 
Martin Corporation, for the U.S. Department of Energy's National  
Nuclear Security Administration under contract DE-AC04-94AL85000.

\end{document}